\documentstyle[12pt,prd,aps,epsfig,preprint]{revtex}
\def\bea{\begin{eqnarray}}

\def\eea{\end{eqnarray}}

\begin{document}

\begin{titlepage}

\title{The effect of supersymmetric CP phases on  Chargino-Pair Production via Drell-Yan Process at the LHC}

\author{Kerem CANKOCAK}

\address{Department of Physics, Mugla University, Mugla- Turkey}

\author{Aytekin AYDEMIR}

\address{Department of Physics, Mersin University, Mersin-Turkey}

\author{Ramazan SEVER}

\address{Department of Physics, Middle East Technical University, Ankara- Turkey}

\date{\today }

\thispagestyle{empty}

\baselineskip=15pt

\thispagestyle{empty}

\maketitle

\begin{abstract}

We compute the rates for $pp$ annihilation into chargino-pairs via
Drell-Yan process taking into account the effects of
supersymmetric soft phases, at proton-proton collider. In
particular, the phase of the $\mu$ parameter gains direct
accessibility via the production of dissimilar charginos. The
phases of the trilinear soft masses do not have a significant
effect on the cross sections.

%Our results can be important for sparticle searches at the LHC.

\end{abstract}


\end{titlepage}

\newpage

\section{Introduction}

Supersymmetry (SUSY), is one of the most favored extensions of
the SM which is capable of stabilizing the ino-sector of
fundamental scalars against the ultraviolet divergences. The
(soft) breaking of SUSY, around the ${\rm TeV}$ scale, brings
about two new ingredients compared to the standard electroweak
theory (SM): First, there are novel sources of flavor violation
coming through the off--diagonal entries of the squark mass
matrices. Second, there are novel sources of CP violation coming
from the phases of the soft masses. The first effect, which cannot
be determined theoretically, is strongly constrained by the FCNC
data \cite{masiero} , and therefore, as a predictive case, it is
convenient to restrict all flavor--violating transitions to the
charged--current interactions where they proceed via the known CKM
angles. However, this very restriction of the flavor violation to
the SM one does not evade new sources of CP violation. Indeed, the
model possesses various flavor--blind CP--odd phases contained in
the complex $\mu $ parameter, $A$ parameters, and gauge fermion
masses $M_{i}$.

These phases form the new sources of CP violation which shows up
in the electric dipole moments (EDMs) of leptons and hadrons (See
\cite{edm1} and references therein). For heavy quark EDMs see
\cite{edm2} and for the rate asymmetries of various heavy--light
mesons \cite{meson}. Therefore, it is of fundamental importance to
determine appropriate collider processes where all or some of the
SUSY CP phases can be inferred or measured. In fact, the effects
of the SUSY CP phases on the Higgs production has already been
analyzed in \cite{higgs,higgsp}. In this work we will discuss the
chargino-pair production at LHC energies and ways of isolating the
phase of the $\mu $ parameter from the cross section.

 Due to the large energy and high luminosity of incoming protons,
 LHC is a very useful machine to detect charginos. The dominant
 production mechanism for chargino pairs at a hadron collider
 is the quark-antiquark annihilation. In this sense, the number of
chargino pair production events differ from $pp$ to
$p\overline{p}$ colliders. In what follows we will compute the
cross section for
$pp\rightarrow\tilde{\chi}_{i}^{+}\tilde{\chi}_{j}^{-} + X$ as a
function of $\varphi _{\mu }=\mbox{Arg}[\mu ]$ for various values
of $|\mu |$ and the SU(2) gaugino mass $M_{2}$. As the analyses of
EDMs \cite{edm1,edm2} make it clear the effects of SUSY CP phases
are expected to grow with lowering soft masses.

\section{$q\bar{q}\rightarrow \tilde{\protect\chi}_{i}^{+}\tilde{\protect\chi}_{j}^{-}$}

Our analysis is similar to that used for the linear collider
processes\cite{lc}. The relevant Feynman diagrams are depicted in
Fig. 1. In what follows we mainly deal with the first two diagrams
since the third one is suppressed by presumably heavy squarks.
Then it is obvious that the amplitude for the process depends
exclusively on the phases in the chargino sector, $i.e$, the phase
of the $\mu $ parameter.

Here we summarize the masses and couplings of the charginos for
completeness (See \cite{phase} for details). The charginos which
are the mass eigenstates of charged gauginos and Higgsinos are
described by a $2\times 2$ mass matrix

\begin{equation}
M_{C}=\left(\begin{array}{cc}
M_{2} & \sqrt{2}M_{W}\cos \beta \\
\sqrt{2}M_{W}\sin \beta & |\mu |e^{i\varphi _{\mu }}
\end{array}
\right)
\end{equation}

where $M_{2}$ is the SU(2) gaugino mass taken to be real
throughout the work. The masses of the charginos as well as their
mixing matrices follow from the bi-unitary transformation

\begin{equation}
C_{R}^{\dagger }M_{C}C_{L}=\mbox{diag}(m_{\chi _{1}},m_{\chi_{2}})
\end{equation}

where $C_{L}$ and $C_{R}$ are $2\times 2$ unitary matrices, and
$m_{\chi_{1}}$, $m_{\chi _{2}}$ are the masses of the charginos
$\chi _{1}$, $\chi_{2}$ such that $m_{\chi _{1}}<m_{\chi _{2}}$.
It is convenient to choose the following explicit parametrization
for the chargino mixing matrices

\begin{eqnarray}
C_{L} &=&\left(\begin{array}{cc}
\cos \theta _{L} & \sin \theta _{L}e^{i\varphi _{L}} \\
-\sin \theta _{L}e^{-i\varphi _{L}} & \cos \theta _{L}
\end{array}
\right) \\
C_{R} &=&\left(
\begin{array}{cc}
\cos \theta _{R} & \sin \theta _{R}e^{i\varphi _{R}} \\
-\sin \theta _{R}e^{-i\varphi _{R}} & \cos \theta _{R}
\end{array}
\right) \cdot \left(
\begin{array}{cc}
e^{i\phi _{1}} & 0 \\
0 & e^{i\phi _{2}}
\end{array}
\right)
\end{eqnarray}
where the angle parameters $\theta _{L,R}$, $\varphi _{L,R}$, and
$\phi_{1,2}$ can be determined from the defining Eq. (1). A
straightforward calculation yields

\begin{eqnarray}
\tan 2\theta _{L} &=&\frac{\sqrt{8}M_{W}\sqrt{M_{2}^{2}\cos
^{2}\beta +|\mu|^{2}\sin ^{2}\beta +|\mu |M_{2}\sin 2\beta \cos
\varphi _{\mu }}}{M_{2}^{2}-|\mu |^{2}-2M_{W}^{2}\cos 2\beta }  \nonumber \\
\tan 2\theta _{R} &=&\frac{\sqrt{8}M_{W}\sqrt{|\mu
|^{2}\cos^{2}\beta+M_{2}^{2}\sin ^{2}\beta +|\mu |M_{2}\sin 2\beta
\cos \varphi _{\mu }}}{M_{2}^{2}-|\mu |^{2}+2M_{W}^{2}\cos 2\beta }  \nonumber \\
\tan \varphi _{L} &=&\frac{|\mu |\sin \varphi _{\mu }}{M_{2}\cot
\beta +|\mu|\cos \varphi _{\mu }}  \nonumber \\
\tan \varphi _{R} &=&-\frac{|\mu |\cot \beta \sin \varphi _{\mu
}}{|\mu|\cot \beta \cos \varphi _{\mu }+M_{2}}
\end{eqnarray}

in terms of which the remaining two angles $\phi _{1}$ and $\phi
_{2}$ read as follows

\begin{equation}
\tan \phi _{i}=\frac{\mbox{Im}[Q_{i}]}{\mbox{Re}[Q_{i}]}
\end{equation}

where $i=1,2$ and

\begin{eqnarray}
Q_{1} &=&\sqrt{2}M_{W}[\cos \beta \sin \theta _{L}\cos
\theta_{R}e^{-i\varphi _{L}}+\sin \beta \cos \theta _{L}
\sin\theta_{R}e^{i\varphi _{R}}]  \nonumber \\
&+&M_{2}\cos \theta _{L}\cos \theta _{R}+|\mu |\sin \theta_{L}\sin
\theta_{R}e^{i(\varphi _{\mu }+\varphi _{R}-\varphi _{L})}  \nonumber \\
Q_{2} &=&-\sqrt{2}M_{W}[\cos \beta \sin \theta _{R}\cos
\theta_{L}e^{-i\varphi _{R}}+\sin \beta \cos \theta _{R}\sin
\theta_{L}e^{i\varphi _{L}}]  \nonumber \\
&+&M_{2}\sin \theta _{L}\sin
\theta_{R}e^{i(\varphi_{L}-\varphi_{R})}+|\mu |\cos \theta_{L}\cos
\theta _{R}e^{i\varphi _{\mu }}\,.
\end{eqnarray}

The origin of the phases $\theta _{L,R}$, $\varphi _{L,R}$, and
$\phi _{1,2}$ is easy to trace back. The angles $\theta _{L}$ and
$\theta _{R}$ would be sufficient to diagonalize, respectively,
the quadratic mass matrices $M_{C}^{\dagger }M_{C}$ and
$M_{C}M_{C}^{\dagger }$ if $M_{C}$ were real. As a result one
needs the additional phases $\varphi _{L,R}$ which are identical
to the phases in the off--diagonal entries of the matrices
$M_{C}^{\dagger }M_{C}$ and $M_{C}M_{C}^{\dagger }$, respectively.
However, these four phases are still not sufficient for making the
chargino masses real positive due to the bi-unitary nature of the
transformation, and hence, the phases $\phi _{1}$ and $\phi _{2}$
can not also be made real positive. Finally, inserting the unitary
matrices $C_{L}$ and $C_{R}$ into the defining Eq. (1) one obtains
the following expressions for the masses of the charginos

\begin{eqnarray}
m_{\chi _{1(2)}}^{2}
&=&\frac{1}{2}\Big\{M_{2}^{2}+|\mu|^{2}+2M_{W}^{2}-(+)[(M_{2}^{2}-|\mu
|^{2})^{2}+4M_{W}^{4}\cos ^{2}2\beta\nonumber \\
&+&4M_{W}^{2}(M_{2}^{2}+|\mu |^{2}+2M_{2}|\mu |\sin 2\beta \cos
\varphi_{\mu })]^{1/2}\Big\}.
\end{eqnarray}

The fundamental SUSY parameters $M_{2}$, $|\mu |$, $\tan \beta $
and the phase parameter $\cos \varphi _{\mu }$ can be extracted
from the chargino $\tilde{\chi}_{1,2}^{\pm }$ parameters \cite{lc}
$i.e.$ the masses $m_{\tilde{\chi}_{1,2}^{\pm }}$ and the two
mixing angles $\phi _{L}$ and $\phi _{R}$ of the left and right
chiral components of the wave function. These mixing angles are
physical observables and they can be measured just like the
chargino masses $m_{\tilde{\chi}_{1,2}^{\pm }}$ in the process
$q+\bar{q}\rightarrow \tilde{\chi}_{i}^{+}+\tilde{\chi}_{j}^{-}$.
The two angles $\phi_{L}$ and $\phi _{R}$ and the nontrivial phase
angles $\{\varphi _{L},$ $ \varphi _{R},$ $\phi _{1},\phi _{2}\}$
define the couplings of the chargino-chargino-Z vertices

\[
\left\langle \tilde{\chi}_{1L}^{-}\left| Z\right|
\tilde{\chi}_{1L}^{-} \right\rangle
=-\frac{e}{s_{W}c_{W}}[s_{W}^{2}-\frac{3}{4}-\frac{1}{4}\cos
2\theta _{L}]
\]

\[
\left\langle \tilde{\chi}_{1L}^{-}\left| Z\right|
\tilde{\chi}_{2L}^{-}\right\rangle=+\frac{e}{4s_{W}c_{W}}e^{-i\varphi_{L}}\sin
2\theta_{L}
\]

\[
\left\langle \tilde{\chi}_{2L}^{-}\left| Z\right|
\tilde{\chi}_{2L}^{-} \right\rangle
=-\frac{e}{s_{W}c_{W}}[s_{W}^{2}-\frac{3}{4}+\frac{1}{4}\cos
2\theta _{L}]
\]

\[
\left\langle \tilde{\chi}_{1R}^{-}\left| Z\right|
\tilde{\chi}_{1R}^{-} \right\rangle
=-\frac{e}{s_{W}c_{W}}[s_{W}^{2}-\frac{3}{4}-\frac{1}{4}\cos
2\theta _{R}]
\]

\[
\left\langle \tilde{\chi}_{1R}^{-}\left| Z\right|
\tilde{\chi}_{2R}^{-}\right\rangle
=+\frac{e}{4s_{W}c_{W}}e^{-i(\varphi _{R}-\phi_{1}+\phi _{2})}\sin
2\theta _{R}
\]

\begin{equation}
\left\langle \tilde{\chi}_{2R}^{-}\left| Z\right|
\tilde{\chi}_{2R}^{-} \right\rangle
=-\frac{e}{s_{W}c_{W}}[s_{W}^{2}-\frac{3}{4}+\frac{1}{4}\cos
2\theta _{R}]
\end{equation}

where $s_{W}=\sin \theta _{W}$ is the weak angle. Note that every
vertex here is an explicit function of $\varphi _{\mu }$ via the
various mixing angles. However, the $Z$ coupling to unlike
charginos $\tilde{\chi}_{i}^{+}\tilde{\chi}_{j}^{-}$ is manifestly
complex, and its phase vanishes in the CP--conserving limit,
$\varphi _{\mu }\rightarrow 0,$ $\pi $.

Obviously, the photon vertex is independent of the SUSY phases

\begin{equation}
\left\langle \tilde{\chi}_{iL,R}^{-}\left| \gamma \right|
\tilde{\chi}_{jL,R}^{-}\right\rangle =e\delta _{ij}
\end{equation}

The process $q\bar{q}\rightarrow
\tilde{\chi}_{i}^{+}\tilde{\chi}_{j}^{-}$ is generated by the two
mechanisms shown in Fig. 1. The $s$-channel $\gamma $ and $Z$
exchanges, and $t$-channel $\tilde{q}$ exchange, where the latter
is consistently neglected below. The transition amplitude can be
parameterized as

\begin{equation}
T(q\bar{q}\rightarrow
\tilde{\chi}_{i}^{+}\tilde{\chi}_{j}^{-})=\frac{e^{2}}{s}Q_{\alpha
\beta } [\bar{v}(\bar{q})\gamma _{\mu }P_{\alpha
}u(q)][\bar{u}(\tilde{\chi}_{i}^{-})\gamma ^{\mu }P_{\beta
}v(\tilde{\chi}_{j}^{+})]
\end{equation}

where the charges $Q_{\alpha \beta }$ are defined such that the
first index corresponds to the chirality of the $\overline{q}q$
current and the second one to chargino current. For various final
states, their expressions are given by

(i)\qquad \underline{$\tilde{\chi}_{1}^{-}\tilde{\chi}_{1}^{+}$}
for \underline{$q=u,c$}

\[
Q_{LL}=1+\frac{D_{Z}}{s_{W}^{2}c_{W}^{2}}(\frac{1}{2}-\frac{2}{3}s_{W}^{2})(s_{W}^{2}-
\frac{3}{4}-\frac{1}{4}\cos 2\phi _{L})
\]

\[
Q_{LR}=1+\frac{D_{Z}}{s_{W}^{2}c_{W}^{2}}(\frac{1}{2}-\frac{2}{3}s_{W}^{2})(s_{W}^{2}-
\frac{3}{4}-\frac{1}{4}\cos2\phi _{R})
\]

\[
Q_{RL}=1+\frac{D_{Z}}{c_{W}^{2}}(-\frac{2}{3})(s_{W}^{2}-\frac{3}{4}-\frac{1}{4}\cos
2\phi _{L})
\]

\begin{equation}
Q_{RR}=1+\frac{D_{Z}}{c_{W}^{2}}(-\frac{2}{3})(s_{W}^{2}-\frac{3}{4}-\frac{1}{4}\cos
2\phi _{R})
\end{equation}

(ii)\qquad \underline{$\tilde{\chi}_{1}^{-}\tilde{\chi}_{1}^{+}$}
for \underline{$q=d,s$}

\[
Q_{LL}=1+\frac{D_{Z}}{s_{W}^{2}c_{W}^{2}}(-\frac{1}{2}+\frac{1}{3}s_{W}^{2})
(s_{W}^{2}-\frac{3}{4}-\frac{1}{4}\cos2\phi _{L})
\]

\[
Q_{LR}=1+\frac{D_{Z}}{s_{W}^{2}c_{W}^{2}}(-\frac{1}{2}+\frac{1}{3}s_{W}^{2})(s_{W}^{2}-
\frac{3}{4}-\frac{1}{4}\cos2\phi _{R})
\]

\[
Q_{RL}=1+\frac{D_{Z}}{c_{W}^{2}}(+\frac{1}{3})(s_{W}^{2}-\frac{3}{4}-\frac{1}{4}\cos
2\phi _{L})
\]

\begin{equation}
Q_{RR}=1+\frac{D_{Z}}{c_{W}^{2}}(+\frac{1}{3})(s_{W}^{2}-\frac{3}{4}-\frac{1}{4}\cos
2\phi _{R})
\end{equation}

\bigskip

(iii)\qquad \underline{$\tilde{\chi}_{1}^{-}\tilde{\chi}_{2}^{+}$}
for \underline{$q=u,c$}

\[
Q_{LL}=\frac{D_{Z}}{4s_{W}^{2}c_{W}^{2}}(\frac{1}{2}-\frac{2}{3}s_{W}^{2})e^{-i\varphi
_{L}}\sin 2\phi _{L}
\]

\[
Q_{LR}=\frac{D_{Z}}{4s_{W}^{2}c_{W}^{2}}(\frac{1}{2}-\frac{2}{3}s_{W}^{2})e^{-i(\varphi
_{R}-\phi _{1}+\phi _{2})}\sin 2\phi _{R}
\]

\[
Q_{RL}=\frac{D_{Z}}{4c_{W}^{2}}(-\frac{2}{3})e^{-i\varphi_{L}}\sin
2\phi_{L}
\]

\begin{equation}
Q_{RR}=\frac{D_{Z}}{4c_{W}^{2}}(-\frac{2}{3})e^{-i(\varphi
_{R}-\phi_{1}+\phi _{2})}\sin 2\phi _{R}
\end{equation}

(iv)\qquad \underline{$\tilde{\chi}_{1}^{-}\tilde{\chi}_{2}^{+}$}
for \underline{$q=d,s$}

\[
Q_{LL}=\frac{D_{Z}}{4s_{W}^{2}c_{W}^{2}}(-\frac{1}{2}+\frac{1}{3}s_{W}^{2})e^{-i\varphi
_{L}}\sin 2\phi _{L}
\]

\[
Q_{LR}=\frac{D_{Z}}{4s_{W}^{2}c_{W}^{2}}(-\frac{1}{2}+\frac{1}{3}s_{W}^{2})e^{-i(\varphi
_{R}-\phi _{1}+\phi _{2})}\sin 2\phi _{R}
\]

\[
Q_{RL}=\frac{D_{Z}}{4c_{W}^{2}}(+\frac{1}{3})e^{-i\varphi
_{L}}\sin 2\phi_{L}
\]

\begin{equation}
Q_{RR}=\frac{D_{Z}}{4c_{W}^{2}}(+\frac{1}{3})e^{-i(\varphi
_{R}-\phi_{1}+\phi _{2})}\sin 2\phi _{R}
\end{equation}

(v)\qquad \underline{$\tilde{\chi}_{2}^{-}\tilde{\chi}_{2}^{+}$}
for \underline{$q=u,c$}

\[
Q_{LL}=1+\frac{D_{Z}}{s_{W}^{2}c_{W}^{2}}(\frac{1}{2}-\frac{2}{3}s_{W}^{2})
(s_{W}^{2}-\frac{3}{4}+\frac{1}{4}\cos2\phi _{L})
\]

\[
Q_{LR}=1+\frac{D_{Z}}{s_{W}^{2}c_{W}^{2}}(\frac{1}{2}-\frac{2}{3}
s_{W}^{2})(s_{W}^{2}-\frac{3}{4}+\frac{1}{4}\cos2\phi _{R})
\]

\[
Q_{RL}=1+\frac{D_{Z}}{c_{W}^{2}}(-\frac{2}{3})(s_{W}^{2}-
\frac{3}{4}+\frac{1}{4}\cos2\phi_{L})
\]

\begin{equation}
Q_{RR}=1+\frac{D_{Z}}{c_{W}^{2}}(-\frac{2}{3})(s_{W}^{2}-\frac{3}{4}+\frac{1}{4}\cos
2\phi _{R})
\end{equation}

(vi)\qquad \underline{$\tilde{\chi}_{2}^{-}\tilde{\chi}_{2}^{+}$}
for \underline{$q=d,s$}

\[
Q_{LL}=1+\frac{D_{Z}}{s_{W}^{2}c_{W}^{2}}(-\frac{1}{2}+\frac{1}{3}
s_{W}^{2})(s_{W}^{2}-\frac{3}{4}+\frac{1}{4}\cos2\phi _{L})
\]

\[
Q_{LR}=1+\frac{D_{Z}}{s_{W}^{2}c_{W}^{2}}(-\frac{1}{2}+\frac{1}{3}
s_{W}^{2})(s_{W}^{2}-\frac{3}{4}+\frac{1}{4}\cos2\phi _{R})
\]

\[
Q_{RL}=1+\frac{D_{Z}}{c_{W}^{2}}(+\frac{1}{3})(s_{W}^{2}-
\frac{3}{4}+\frac{1}{4}\cos2\phi _{L})
\]

\begin{equation}
Q_{RR}=1+\frac{D_{Z}}{c_{W}^{2}}(+\frac{1}{3})(s_{W}^{2}-\frac{3}{4}+\frac{1}{4}\cos
2\phi _{R})
\end{equation}

Here all the amplitudes are built up by the $\gamma$ and $Z$
exchanges, and $D(Z)$ stands for the $Z$ propagator:
$D_{Z}=\hat{s}/(\hat{s}-m_{Z}^{2}+im_{Z}\Gamma_{Z}) $.

In what follows, for convenience we will introduce four
combinations of the charges

\begin{eqnarray}
Q_{1} &=&\frac{1}{4}[\left| Q_{RR}\right| ^{2}+\left|
Q_{LL}\right|^{2}+\left| Q_{RL}\right| ^{2}+\left| Q_{LR}\right| ^{2}]  \nonumber \\
Q_{2} &=&\frac{1}{2}Re[Q_{RR}Q_{RL}^{\ast }+Q_{LL}Q_{LR}^{\ast }]
\nonumber
\\
Q_{3} &=&\frac{1}{4}[\left| Q_{RR}\right| ^{2}+\left|
Q_{LL}\right|^{2}-\left| Q_{RL}\right| ^{2}-\left| Q_{LR}\right| ^{2}]  \nonumber \\
Q_{4} &=&\frac{1}{2}Im[Q_{RR}Q_{RL}^{\ast }+Q_{LL}Q_{LR}^{\ast }]
\end{eqnarray}

so that the differential cross section can be expressed simply as

\begin{equation}
\frac{d\hat{\sigma} }{d\cos \Theta }(q\bar{q}\rightarrow
\tilde{\chi}_{i}^{+}\tilde{\chi}_{j}^{-})=\frac{\pi \alpha
^{2}}{2\hat{s}}\lambda ^{1/2}\{[1-(\mu_{i}^{2}-\mu
_{j}^{2})^{2}+\lambda \cos ^{2}\Theta ]Q_{1}+4\mu
_{i}\mu_{j}Q_{2}+2\lambda ^{1/2}Q_{3}\cos \Theta \}
\end{equation}

with the usual two body phase space factor

\begin{equation}
\lambda (1,\mu _{i}^{2},\mu _{j}^{2})=[1-(\mu
_{i}+\mu_{j})^{2}][1-(\mu_{i}-\mu _{j})^{2}]
\end{equation}

defined via the reduced mass $\mu
_{i}^{2}=\frac{m_{\tilde{\chi}_{i}^{\pm}}^{2}}{\hat{s}}$.

Integrating the differential cross section over the
center--of--mass scattering angle $\Theta $ one can obtain the
total partonic cross section

\begin{equation}
\hat{\sigma} =\hat{\sigma} (\varphi _{\mu },\mu
,M_{2},\hat{s},\tan \beta )
\end{equation}

whose dependence on $\varphi _{\mu },M_{2}$ and $|\mu |$ will be
analyzed numerically.

The total cross section for the chargino-pair production through
$q\bar{q}$ fusion  at the LHC center of mass energy $\sqrt{s}=14$
TeV, can be obtained by integrating partonic cross section
$\hat{\sigma}$, over the quark-antiquark luminosities in the
distribution function of proton, which is given in parton model by
the formula\cite{refparton}

\begin{equation}
 \sigma(pp\rightarrow{q\bar{q}}\rightarrow
 \tilde{\chi}_{i}^{+}\tilde{\chi}_{j}^{-} + X) =
\int^1_{\frac{Q^2}{s}} {d\tau} \hskip 1mm {\frac{d{\cal
L}_{q\bar{q}}}{d\tau}} \hskip 2mm
{\hat{\sigma}}_{q\bar{q}}(q\bar{q}\rightarrow
\tilde{\chi}_{i}^{+}\tilde{\chi}_{j}^{-}, \hskip 3mm at \hskip 3mm
\hat{s}=\tau s )
\end{equation}

with a quark luminosity in proton which is defined as

\begin{equation}
 {\frac{d{\cal L}_{q\bar{q}}}{d\tau}}=\int^1_{\tau} {\frac{dx_A}{x_A}}
 {\sum_{ij=q,\bar{q}}\ f_{i/A}(x_A,Q^2) f_{j/B}(x_B,Q^2) }
\end{equation}

 where $f_{i/A}(x_A,Q^2)$ are the
parton distribution functions for parton $i$ in hadron $A$ with
momentum fraction $x_A$ evaluated at the factorization scale
$Q^2=(m_{{\chi}_{i}^{+}} + m_{{\chi}_{j}^{+}})^2$, and \\
$\tau=x_{A}x_{B}$. Here, $s$ is the hadron-hadron center of mass
energy squared which is related to $\hat{s}$, the parton-parton
center of mass energy squared, via $\hat{s}=x_A x_B s$.

To calculate the total cross sections for LHC, one has to know
parton distributions as functions of the scaling variables
$x_{A,B}$ and $Q^2$. Although the distributions have not been
measured at such values of $Q^2$, one can obtain them using
Altarelli-Parisi equation\cite{ap} and the parton distributions at
some scale $Q_{0}^{2}$. For this purpose, we used CTEQ4M parton
densities, without any QCD corrections for simplicity in our
analysis\cite{cteq}.

These results are illustrated in Figs. 2-4, where the production
cross-section is plotted against the phase angle  at the LHC with
$ \sqrt{s}=14$ TeV.

Besides these, it is necessary to analyze the rate asymmetries for
having a better information about $\varphi_{\mu }$.

Analyzing the effects due to normal polarization of the charginos,
the CP-violating phase $\varphi_{\mu }$ can be determined.
Concerning this point, we investigate the normal polarization
vector of the charginos which are inherently CP--odd and therefore
exist if CP is broken in the fundamental theory. The normal
polarization vector is defined as

\begin{equation}
P_{N}=8\lambda ^{1/2}\mu _{j}\sin \Theta \frac{Q_{4}}{N}
\end{equation}

for $\tilde{\chi}_{j}^{+}\tilde{\chi}_{j}^{-},$ the $j$--th
chargino, and it is defined as

\begin{eqnarray}
P_{N}[\tilde{\chi}_{i,j}^{\pm }] &=&\pm 4\lambda
^{1/2}\mu_{j,i}(F_{R}^{2}-F_{L}^{2})\sin \Theta \sin 2\phi _{L}
\nonumber
\\ &&\times \sin 2\phi _{R}\sin (\beta _{L}-\beta _{R}+\gamma_{1}-\gamma _{2})/N
\end{eqnarray}

for non-diagonal pairs $\tilde{\chi}_{i}^{+}\tilde{\chi}_{j}^{-}$
where $i\neq j$. Here

\begin{equation}
N=4[(1-(\mu _{i}^{2}-\mu _{j}^{2})^{2}+\lambda \cos
^{2}\Theta)Q_{1}+4\mu _{i}\mu _{j}Q_{2}+2\lambda ^{1/2}Q_{3}\cos
\Theta ]
\end{equation}

and

\begin{equation}
F_{R}=\frac{D_{Z}}{4c_{W}^{2}},\text{
}F_{L}=\frac{D_{Z}}{4s_{W}^{2}c_{W}^{2}}(s_{W}^{2}-\frac{1}{2}).
\end{equation}

A non-vanishing $P_{N}$ will be sufficient to establish
non-vanishing CP violation in the system. Therefore, the value of
non-vanishing $P_{N}$ implies the strength of the CP invariance
breaking in SUSY.

%\newpage **************************************************

\section{Numerical Estimates}

%\subsection{Chargino production}

In this section we will discuss the dependence of the chargino
production cross section on $\varphi _{\mu },M_{2}$ and $|\mu |$
at $\sqrt{s}=14$ TeV. We apply everywhere the existing collider
constraint that $m_{\chi_{2}}>104\ GeV$.

In Table 1, we give the cross section values for $pp\rightarrow
\tilde{\chi}_{1}^{+}\tilde{\chi}_{1}^{-} + X $ and $pp\rightarrow
\tilde{\chi}_{1}^{+}\tilde{\chi}_{2}^{-}  + X $ taking
M$_{2}=150,$ $300$ \ GeV, $\mu=150,300$ GeV, $\tan \beta =4,$
$10,$ $30,$ $50,$ and $\varphi _{\mu}=\pi /3$  in the
calculations.

In Figs. 2 and 3 we show the dependence of the cross sections
$pp\rightarrow \tilde{\chi}_{1}^{+}\tilde{\chi}_{1}^{-}+ X$ and
$pp\rightarrow \tilde{\chi}_{1}^{+}\tilde{\chi}_{2}^{-}+ X$ on
$\varphi _{\mu }$, \ for M$_{2}=150,$ $300$ \ GeV, $\mu=150,300$
GeV, and $\tan \beta =10.$

The variation of the cross section makes it clear that, as
$\varphi_{\mu }$ varies from 0 to $\pi $ the cross section
decreases gradually. The more spectacular enhancement implies the
lighter chargino mass.

The decrease of the cross section is tied up to the variation of
the chargino masses with the phases. It is clear that as
$\varphi_{\mu}:0\rightarrow \pi $ the mass splitting of the
charginos decrease. This is an important effect which implies that
the cross section is larger than what one would expect from the
CP--conserving theory \cite{lc}.

Apart from the cross section itself, one can analyze various spin
and charge asymmetries which are expected to have an enhanced
dependence on $\varphi_{\mu }$. The normal polarization in
$q\bar{q}\rightarrow \tilde{\chi}_{1}^{+}\tilde{\chi}_{1}^{-}$ is
zero since the $\tilde{\chi}_{1}^{+}\tilde{\chi}_{1}^{-}\gamma $
and $\tilde{\chi}_{1}^{+}\tilde{\chi}_{1}^{-}Z$ vertices are real
even for non-zero phases in the chargino mass matrix.

In Fig. 6 we show the normal polarization $P_{N}$ of unlike
charginos in $q\bar{q}\rightarrow
\tilde{\chi}_{1}^{+}\tilde{\chi}_{2}^{-}$\ which has a different
dependence on the phases. Here again M$_{2}=150,300$\ GeV, $\mu
=150,300$ GeV, $\tan \beta =10,$ and  $\varphi_{\mu }=\pi/2.$ The
value of $\varphi_{\mu } $ is chosen since the normal polarization
is maximum at $\varphi_{\mu }=\pi/2 $.  This result will be
discussed in the next section.

The dependence of the normal polarization on the value of $\Theta$
and $\varphi _{\mu } $ \ is shown in Fig. 7, where the normal
polarization has its maximum at $\Theta = \pi/2$  as expected from
the Equation 24,  and at $\varphi _{\mu }=\pi/2$ as stated above.

We believe that for clarifying the essence of measuring $\varphi
_{\mu }$ the first quantity to be tested is the cross section
itself.

\section{Discussion and Conclusion}

We have analyzed the production of chargino pairs at LHC energies
at the aim of isolating the phase of the $\mu $ parameter. Our
results (Figs. 2-3) suggest that, there is a strong dependence on
the phase of the $\mu $ parameter. The cross section is minimum
for the value of the phase $\varphi _{\mu } = \pi$. This result is
understandable since the mass of the lightest (heaviest) chargino
makes a maximum (minimum) at this point. It is particularly clear
that $\chi_1^+ \chi_1^-$ production rate is depleted in this
region. The size and dependence on $\varphi_{\mu}$ of the cross
section both get enhanced for small enough soft masses, $i.e.$
$M_{2}$=150 GeV and $\mu$=150 GeV. As an example, this is seen for
the value of $\varphi _{\mu } = \pi$, $m_{\chi_{1}}=111\ GeV$ when
$M_{2}$=150 GeV and $\mu$=150 GeV and for the same value of
$\varphi _{\mu }$, $m_{\chi_{1}}=142\ GeV$ when $M_{2}$=150 GeV
and $\mu$=300 GeV. The physical chargino mass increases as the
values of $M_{2}$ and $\mu$ increase.

 The cross section values that we have obtained for $pp\rightarrow
\tilde{\chi}_{1}^{+}\tilde{\chi}_{1}^{-} + X $ process at the LHC
can reach a few 10 $fb$, whereas for the
$\tilde{\chi}_{1}^{+}\tilde{\chi}_{2}^{-}$ pair productions, the
cross sections are in the range of a few $fb$. Having an annual
luminosity of 100 $fb^{-1}$, one may accumulate $10^{3}$ and 100
events per year for $\tilde{\chi}_{1}^{+}\tilde{\chi}_{1}^{-}$ and
$\tilde{\chi}_{1}^{+}\tilde{\chi}_{2}^{-}$ pair productions
respectively. This is a hard task to obtain a clean signature, but
the measurement of these processes will be an important step for
determining the CP violation sources of low--energy supersymmetry.

 The cross sections of the subprocess  ~$q\bar{q}\rightarrow
\tilde{\protect\chi}_{1}^{+}\tilde{\protect\chi}_{1}^{-}$ ~and
~$q\bar{q}\rightarrow
\tilde{\protect\chi}_{1}^{+}\tilde{\protect\chi}_{2}^{-}$ as a
function of ~$q\bar{q}$ ~c.m.s. energy ~$\sqrt{\hat{s}}$ are
depicted in Figs. 4 and 5. There are sharp rising peaks around
~$\sqrt{\hat{s}}$ $\sim$ 300 GeV and $\sqrt{\hat{s}}$ $\sim$ 500
GeV for~$\tilde{\protect\chi}_{1}^{+}\tilde{\protect\chi}_{1}^{-}$
and ~$\tilde{\protect\chi}_{1}^{+}\tilde{\protect\chi}_{2}^{-}$
respectively, due to the threshold conditions ~$\sqrt{\hat{s}}$
$\sim$ $m_{\tilde{\chi}_{i}^{\pm}}+m_{\tilde{\chi}_{j}^{\pm}}$
(i,j=1,2).

Since the cross sections for  ~$q\bar{q}\rightarrow
\tilde{\protect\chi}_{i}^{+}\tilde{\protect\chi}_{j}^{-}$ have
their peaks at the energies less then 1 TeV,  the upgraded
Tevatron as well, with a c.m.s energy of 1.8 TeV and 30 $fb^-1$
integrated luminosity,  is a useful machine to detect CP violation
of SUSY particles. In fact, it is more likely to see chargino
pairs at such $p\overline{p}$ colliders than at the LHC. This
follows from the fact that the anti-quarks needed for chargino
pair production are always drained from the sea for $pp$
colliders. In this sense our analysis puts a lower bound on the
likelihood of observing chargino pairs at hadron colliders. In
this respect, the proton-antiproton collider Tevatron and, to a
greater extent, the proton-proton collider LHC, are the almost
ideal places of probing such a minimal SUSY scenario of explicit
CP violation.

In Fig. 6,  we show the dependence of the normal polarization of
$q\bar{q}\rightarrow \tilde{\chi}_{1}^{+}\tilde{\chi}_{2}^{-}$ on
$\Theta$, \ for M$_{2}=150,$ $300$ \ GeV, $\mu=150,300$ GeV, and
$\tan \beta =10.$ As seen from the figure, $P_{N}$ is not
vanishing for masses of lighter charginos, and this is an
indication of the CP invariance breaking in SUSY. For the higher
values of M$_{2}$ \ and $\mu$, \ $P_{N}$ is practically vanishing
as the mass of charginos increases.

In true experimental environment, the cross sections we have
studied are for subprocess obtained by integration  over
appropriate structure functions. However, given the energy span of
LHC that it will be possible to probe sparticles up to $2\ {\rm
TeV}$, it is clear that the center--of--mass energies we discuss
are always within the experimental reach. If the experiment
concludes $\varphi_{\mu}\sim {\cal{O}}(1)$ then, given strong
bounds from the absence of permanent EDMs for electron, neutron,
atoms and molecules, one would conclude that the first two
generations of sfermions will be hierarchically split from the
ones in the third generation. In case the experiment reports a
small $\varphi_{\mu}$ then presumably all sfermion generations can
lie right at the weak scale in agreement with the EDM bounds. In
this case, where $\varphi_{\mu}$ is a small fraction of $\pi$, one
might expect that the minimal model is UV--completed above the TeV
scale such that the $\mu$ parameter is promoted to a dynamical
SM--singlet field ($e.g.$ the $Z^{\prime}$ models).

\newpage

FIGURE CAPTIONS

FIGURE 1. The lowest order Feynman diagrams for
$q\bar{q}\rightarrow \tilde{\chi}_{i}^{+}\tilde{\chi}_{j}^{-}$
processes.

FIGURE 2. The plot of cross section for $pp\rightarrow
\tilde{\chi}_{1}^{+}\tilde{\chi}_{1}^{-} + X $ as a function of
$\varphi _{\mu }$ for the values of {\bf \ }$\mu =150,300$ GeV,
M$_{2}=150,300$ GeV and $\tan \beta =10$.

FIGURE 3. The plot of cross section for $pp\rightarrow
\tilde{\chi}_{1}^{+}\tilde{\chi}_{2}^{-} + X$ as a function of
$\varphi _{\mu }$ for the values of {\bf \ }$\mu =150,300$ GeV,
 M$_{2}=150,300$ GeV, and $\tan \beta =10$.

FIGURE 4. The plot of cross section for $q\bar{q}\rightarrow
\tilde{\protect\chi}_{1}^{+}\tilde{\protect\chi}_{1}^{-}$  versus
the c.m.s. energy of incoming quarks $\sqrt{\hat{s}}$ for the
values of  $\protect\varphi _{\protect\mu } = 0 $,{\bf \ }$\mu
=150,300$ GeV, M$_{2}=150,300$ GeV and $\tan \beta =10$.

FIGURE 5. The plot of cross section for $q\bar{q}\rightarrow
\tilde{\protect\chi}_{1}^{+}\tilde{\protect\chi}_{2}^{-}$  versus
the c.m.s. energy of incoming quarks $\sqrt{\hat{s}}$ for the
values of $\protect\varphi _{\protect\mu } = 0 $, {\bf \ }$\mu
=150,300$ GeV, M$_{2}=150,300$ GeV and $\tan \beta =10$.

 FIGURES 6. The plot of normal polarization for $q\bar{q}\rightarrow
\tilde{\chi}_{1}^{+}\tilde{\chi}_{2}^{-} $ as a function of
$\Theta $ for the values of $\mu =150,300$ GeV, M$_{2}=150,300$
GeV and $\tan \beta =10$, when $\varphi _{\mu }$ = $\pi /2$ .

 FIGURES 7. 3-dimensional plot of normal polarization for $q\bar{q}\rightarrow
\tilde{\chi}_{1}^{+}\tilde{\chi}_{2}^{-} $ as a function of
$\Theta $ and $\varphi _{\mu }$ for the values of $\mu =150$ GeV,
M$_{2}=150$ GeV and $\tan \beta =10$.

\newpage

\begin{table}[ht]

\caption{ The cross section values for $pp\rightarrow
\tilde{\protect\chi}_{1}^{+}\tilde{\protect\chi}_{1}^{-} + X $ and
$pp\rightarrow
\tilde{\protect\chi}_{1}^{+}\tilde{\protect\chi}_{2}^{-} + X $
processes for $\protect\varphi _{\protect\mu }=\protect\pi /3$,
$\protect\mu =150,300$ GeV, M$_{2}=150,$ $300$ GeV, and
tan$\protect\beta=4,10,30,50.$} \label{Table I.}\unitlength1mm
\centering

\par

\begin{tabular}{|l|l|l|l|l|}

tan$\beta $ & M$_{2}(GeV)$ & $\mu(GeV) $ & $\sigma (pp\rightarrow

\tilde{\chi}_{1}^{+}\tilde{\chi}_{1}^{-}+ X)$(pb) & $\sigma (pp%

\rightarrow \tilde{\chi}_{1}^{+}\tilde{\chi}_{2}^{-}+ X)$(fb) \\
\hline

4 & 150 & 150 & 0.071 & 2.17 \\ \hline

4 & 150 & 300 & 0.056 & 0.50 \\ \hline

4 & 300 & 150 & 0.032 & 0.50 \\ \hline

4 & 300 & 300 & 0.015 & 0.67 \\ \hline

10 & 150 & 150 & 0.067 & 2.09 \\ \hline

10 & 150 & 300 & 0.054 & 0.46 \\ \hline

10 & 300 & 150 & 0.031 & 0.46 \\ \hline

10 & 300 & 300 & 0.014 & 0.66 \\ \hline

30 & 150 & 150 & 0.065 & 2.06 \\ \hline

30 & 150 & 300 & 0.053 & 0.45 \\ \hline

30 & 300 & 150 & 0.03 & 0.45 \\ \hline

30 & 300 & 300 & 0.014 & 0.66 \\ \hline

50 & 150 & 150 & 0.064 & 2.05 \\ \hline

50 & 150 & 300 & 0.052 & 0.45 \\ \hline

50 & 300 & 150 & 0.030 & 0.45 \\ \hline

50 & 300 & 300 & 0.013 & 0.66 \\ \hline
\end{tabular}

\end{table}

\newpage

\begin{figure}[tbph]

\caption{{The lowest order Feynman diagrams for
$q\bar{q}\rightarrow \tilde{\chi}_{i}^{+}\tilde{\chi}_{j}^{-}$
processes. }}

\begin{picture}(161,265)

\put(5,-5){\psfig{file=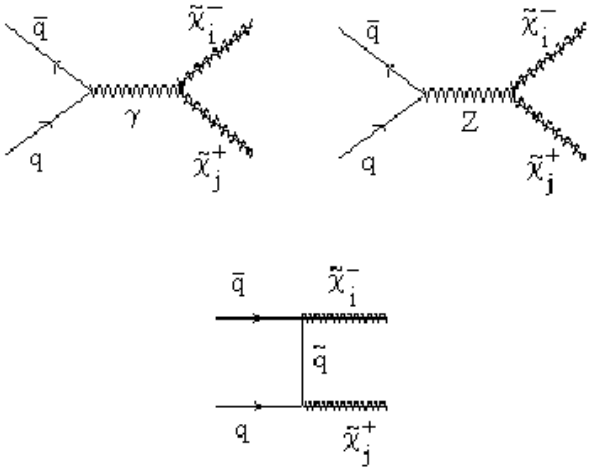,width=10cm} }

\end{picture}

\end{figure}

\newpage

\begin{figure}[tbph]

\caption{{The plot of cross section for $pp\rightarrow
\tilde{\protect\chi}_{1}^{+}\tilde{\protect\chi}_{1}^{-} + X$ as a
function of $\protect\varphi_{\protect\mu }$ for the values of
{\bf \ }$\protect\mu =150,300$ GeV, M$_{2}=150,300$ GeV and $\tan
\beta =10.$ }}

\par

%\vline=10cm.

\begin{picture}(161,265)
\put(-5,-200){\psfig{file=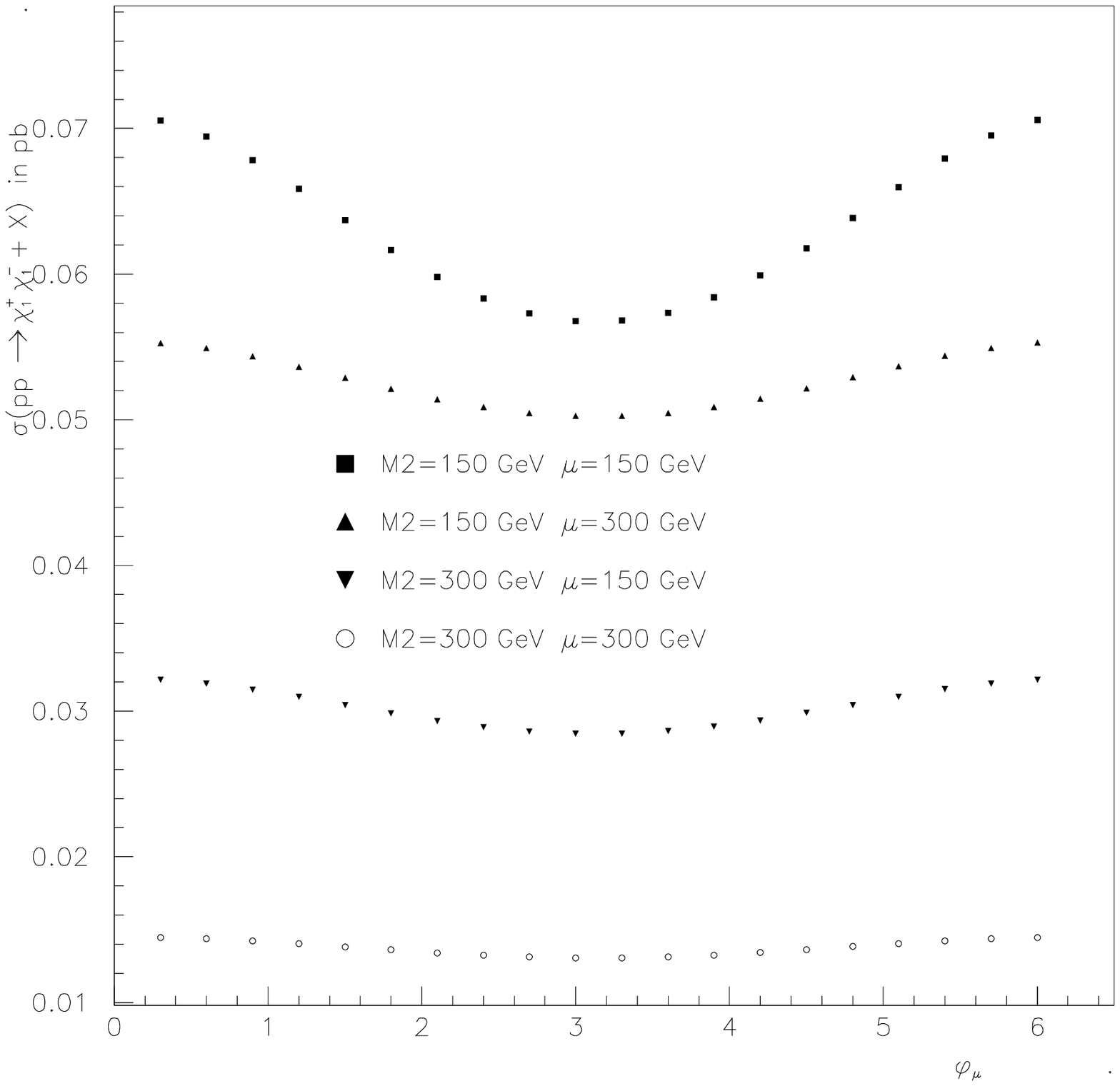,width=17cm} }
%\put(-45,-75){\psfig{file=plot-11.ps} }
\end{picture}
\label{fig2a}
\end{figure}

\newpage

\begin{figure}[tbph]

\caption{{The plot of cross section for $pp\rightarrow
\tilde{\protect\chi}_{1}^{+}\tilde{\protect\chi}_{2}^{-} + X$ as a
function of $\protect\varphi_{\protect\mu }$ for the values of
$\protect\mu =150,300$ GeV, M$_{2}=150,300$ GeV and $\tan \beta
=10.$ }}

\par

\begin{picture}(161,265)
\put(-5,-200){\psfig{file=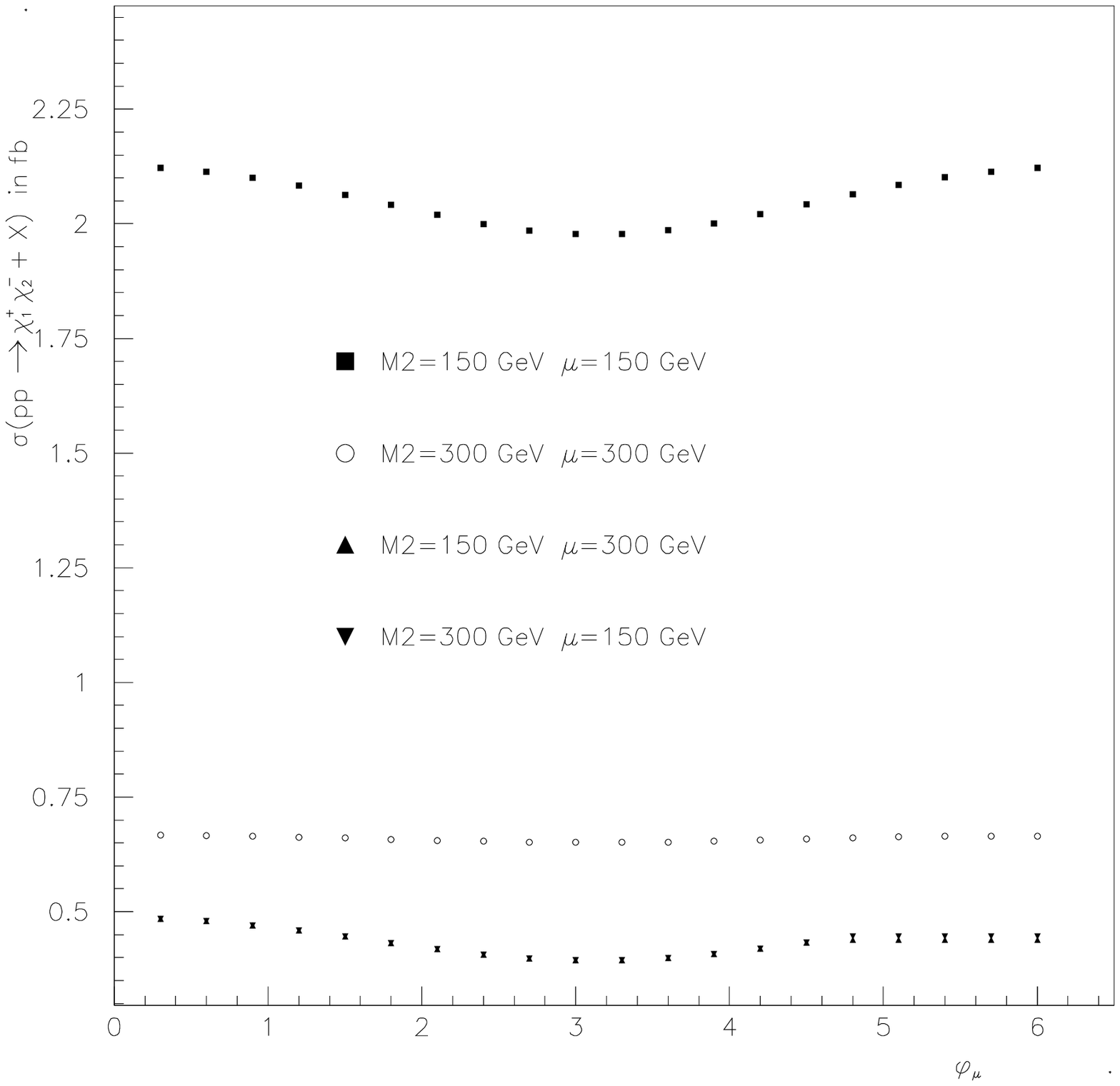,width=17cm} }
%\put(-45,-75){\psfig{file=plot-12.ps} }
\end{picture}
\label{fig3a}
\end{figure}

\newpage

\begin{figure}[tbph]

\caption{{The cross sections of the subprocess
$q\bar{q}\rightarrow
\tilde{\protect\chi}_{1}^{+}\tilde{\protect\chi}_{1}^{-}$  versus
the c.m.s. energy of incoming quarks $\sqrt{\hat{s}}$ for the
values of $\protect\varphi _{\protect\mu } = 0$,  $ \protect\mu
=150,300$ GeV, M$_{2}=150,300$ GeV and $\tan \beta =10.$ }}

\par

\begin{picture}(161,265)
\put(-5,-200){\psfig{file=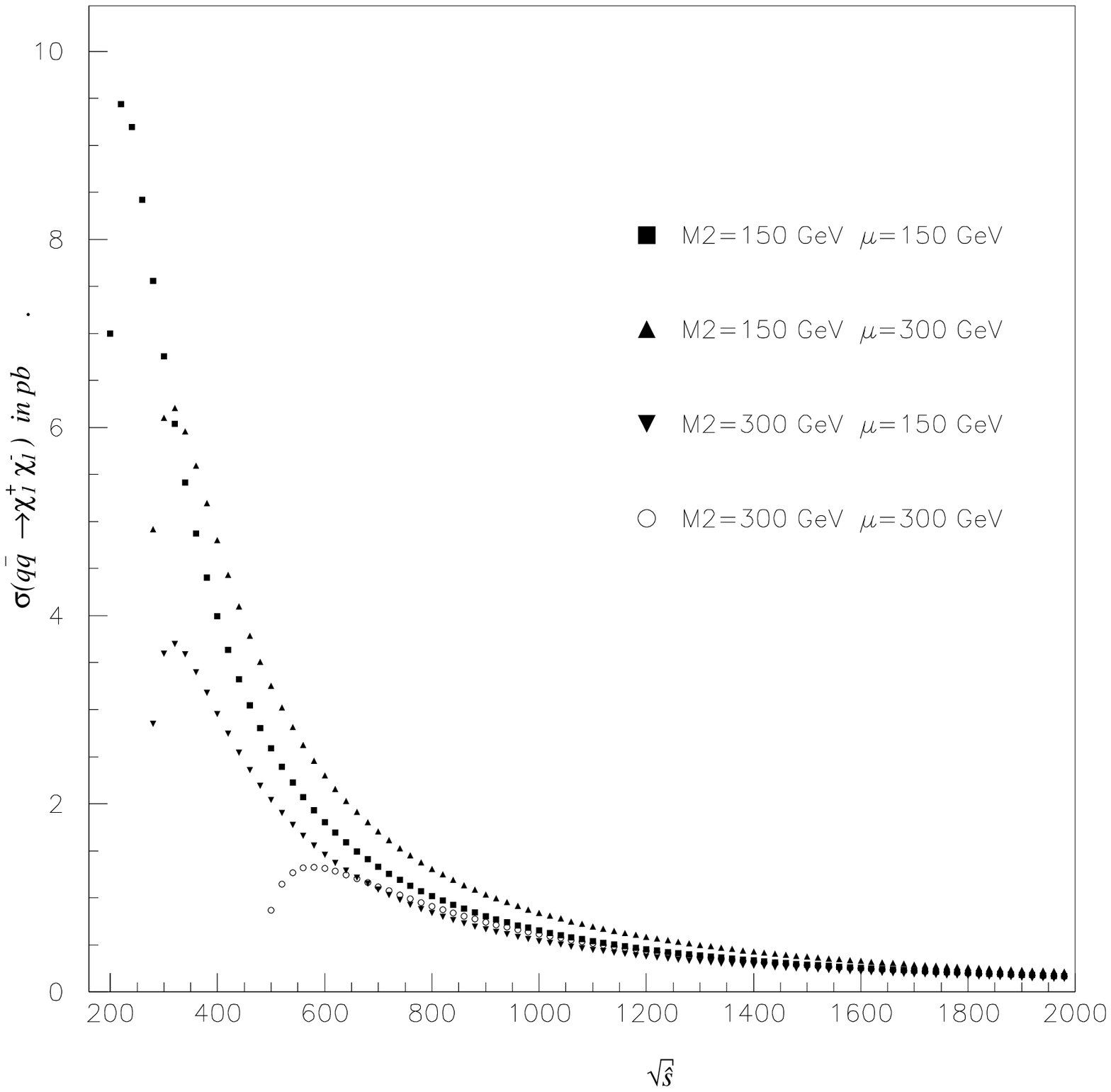,width=17cm} }
\end{picture}
\label{fig4a}
\end{figure}

\newpage

\begin{figure}[tbph]

\caption{{The cross sections of the subprocess
$q\bar{q}\rightarrow
\tilde{\protect\chi}_{1}^{+}\tilde{\protect\chi}_{2}^{-}$  versus
the c.m.s. energy of incoming quarks $\sqrt{\hat{s}}$ for the
values of  $\protect\varphi _{\protect\mu } = 0 $,$\protect\mu
=150,300$ GeV, M$_{2}=150,300$ GeV and $\tan \beta =10.$ }}

\par

\begin{picture}(161,265)
\put(-5,-200){\psfig{file=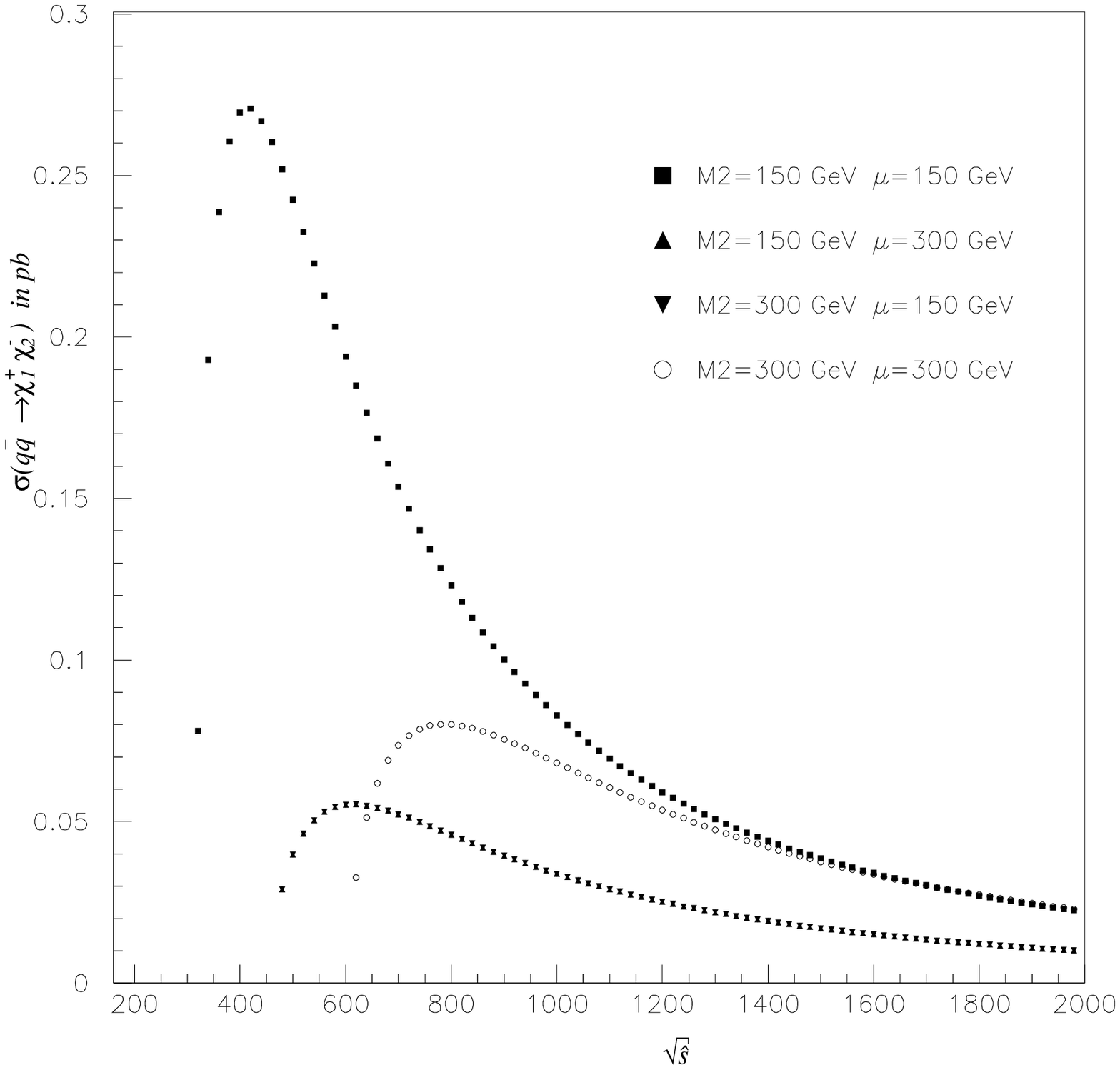,width=17cm} }
\end{picture}
\label{fig4a}
\end{figure}

\newpage

\begin{figure}[tbph]

\caption{{The plot of normal polarization for $q\bar{q}\rightarrow
\tilde{\protect\chi}_{1}^{+}\tilde{\protect\chi}_{2}^{-}$ as a
function of $\Theta$ for the values of $\protect\mu =150,300$ GeV,
 M$_{2}=150,300$ GeV,  $\tan \beta =10$ and $\protect\varphi _{\protect\mu }$ =$\protect\pi /2$ . }}

\par

\begin{picture}(161,265)
\put(-5,-200){\psfig{file=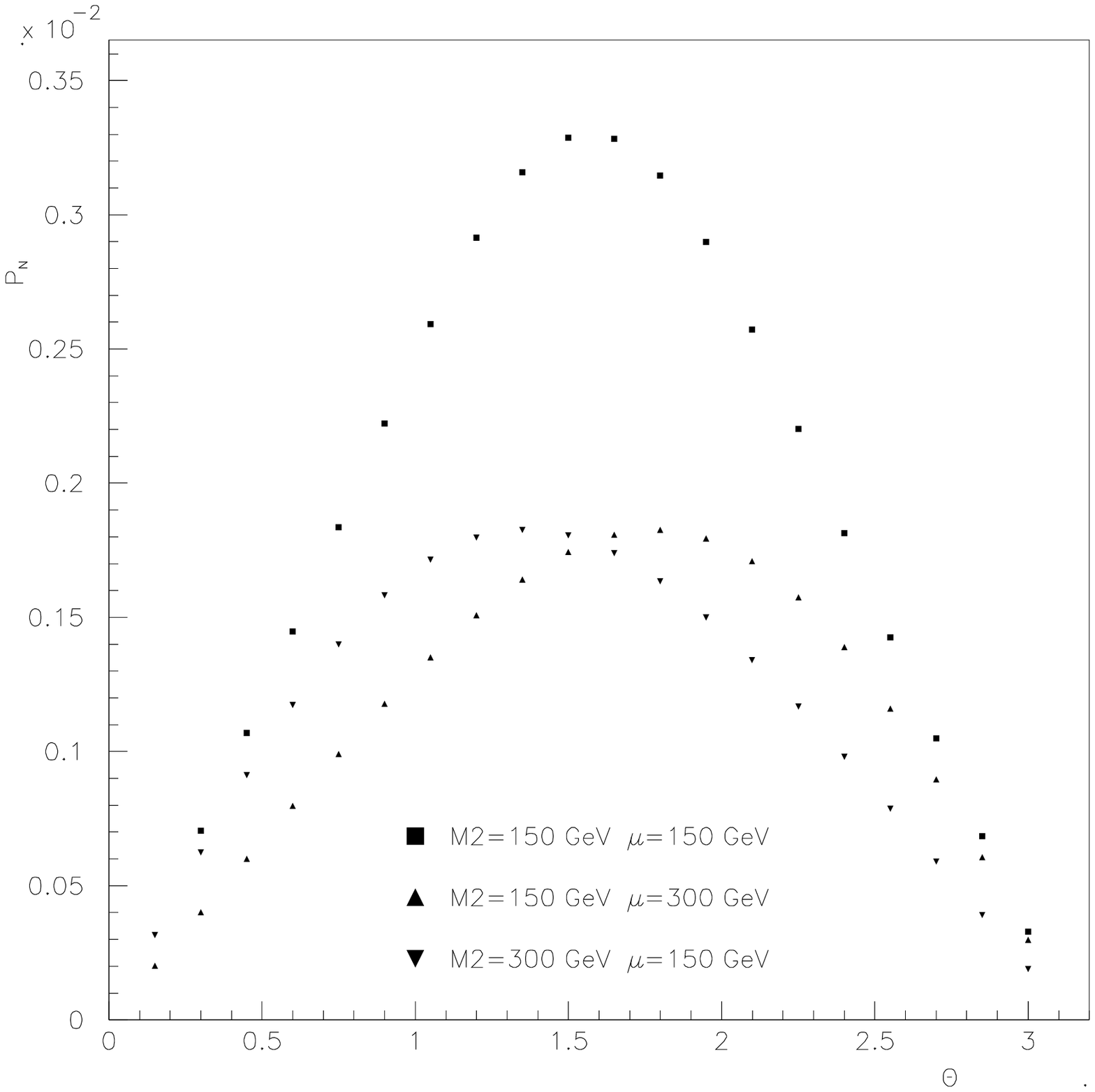,width=17cm} }
\end{picture}
\label{fig4a}
\end{figure}

\newpage

\begin{figure}[tbph]

\caption{{The plot of normal polarization for $q\bar{q}\rightarrow
\tilde{\protect\chi}_{1}^{+}\tilde{\protect\chi}_{2}^{-}$ as a
function of $\Theta$ and  $\protect\varphi _{\protect\mu }$ for
the values of $\protect\mu =150$ GeV,
 M$_{2}=150$ GeV  and $\tan \beta =10.$ }}

\par

\begin{picture}(161,265)
\put(25,-200){\psfig{file=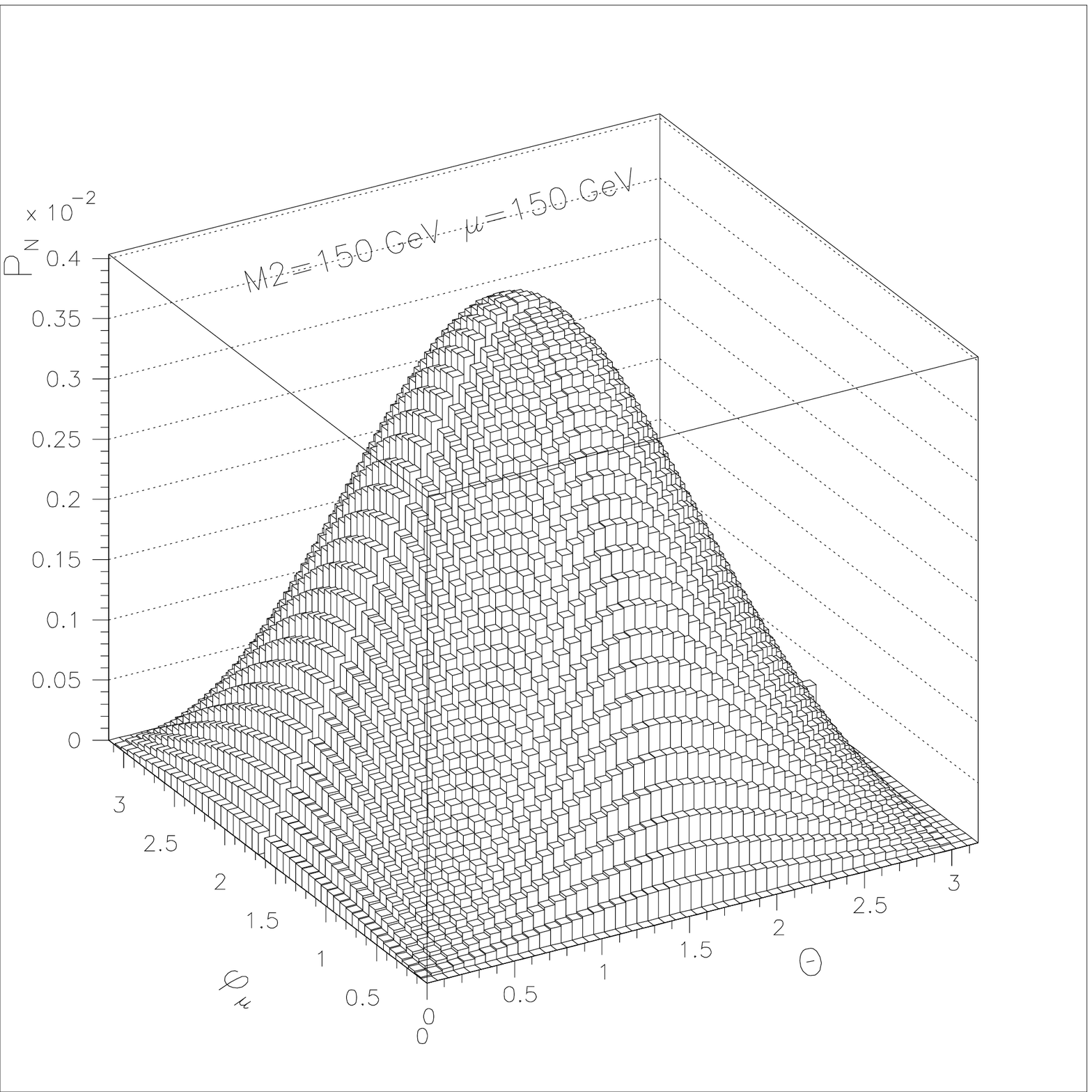,width=14cm} }
\end{picture}
\label{fig4a}
\end{figure}

\end{document}